\documentstyle[12pt,psfig]{article}

\newcommand{\rf}[1]{(\ref{#1})}
\newcommand{\bea}{\begin{eqnarray}}
\newcommand{\eea}{\end{eqnarray}}

\newcommand{\e}{{\rm e}}

\newcommand{\g}{\gamma}

\renewcommand{\L}{\Lambda}
\renewcommand{\b}{\beta}

\newcommand{\n}{\nu}
\newcommand{\m}{\mu}

\newcommand{\ep}{\varepsilon}

\newcommand{\sg}{\sigma}
\newcommand{\k}{\kappa}

\newcommand{\oh}{\frac{1}{2}}

\newcommand{\prt}{\partial}

\newcommand{\cL}{{\cal L}}

\newcommand{\tdqg}{two-dimensional quantum gravity}

\def\void{}
\def\labelmark{}

\newenvironment{formula}[1]{\def\labelname{#1}
\ifx\void\labelname\def\junk{\begin{displaymath}}
\else\def\junk{\begin{equation}\label{\labelname}}\fi\junk}%
{\ifx\void\labelname\def\junk{\end{displaymath}}
\else\def\junk{\end{equation}}\fi\junk\labelmark\def\labelname{}}

{\ifx\void\labelname\def\junk{\end{array}\end{displaymath}}
\else\def\junk{\end{array}\right.\end{equation}}
\fi\junk\labelmark\def\labelname{}\def\junk{}
}

\newcommand{\beq}{\begin{formula}}
\newcommand{\eeq}{\end{formula}}
\newcommand{\beqv}{\begin{formula}{}}

\begin{document}
\topmargin 0pt
\oddsidemargin 5mm
\headheight 0pt
\headsep 0pt
\topskip 9mm

\hfill    NBI-HE-97-53

\hfill January  1998

\begin{center}
\vspace{24pt}
{ \large \bf The Concept of Time in 2D Gravity}

\vspace{24pt}

{\sl J. Ambj\o rn}$\,$\footnote{E-mail: ambjorn@nbi.dk} ,
{\sl K. N. Anagnostopoulos}$\,$\footnote{E-mail: konstant@nbi.dk}, \\
{\sl J. Jurkiewicz}$\,$\footnote{E-mail: jurkiewi@nbi.dk,
~~permanent address:\\ 
~~~~Institute of Physics, Jagellonian University, Reymonta 4, 30-059 Krakow, 
Poland,\\ 
~~~~supported partly by the KBN grants 2P03B04412 and 2P03B19609} 
and 
{\sl C.F. Kristjansen}$\,$\footnote{E-mail: kristjan@nbi.dk}

\vspace{12pt}

The Niels Bohr Institute\\
Blegdamsvej 17, \\
DK-2100 Copenhagen \O , Denmark

\end{center}
\vspace{24pt}

\vfill

\begin{center}
{\bf Abstract}
\end{center}

\vspace{12pt}

\noindent
We show that the ``time'' $t_s$ defined via spin clusters in the 
Ising model coupled to 2d gravity leads to a fractal dimension 
$d_h(s) = 6$ of space-time at the critical point, 
as advocated by Ishibashi and Kawai. In the unmagnetized phase, however,
this definition of Hausdorff dimension breaks down. 
Numerical measurements are consistent with these results. 
The same definition leads to $d_h(s)=16$ at the critical point
when applied to flat space. The fractal dimension $d_h(s)$  is in disagreement 
with both analytical prediction and numerical determination of the fractal 
dimension $d_h(g)$, which is based on the use of the
geodesic distance $t_g$ as ``proper time''. There seems to
be no simple relation of the kind $t_s = t_g^{d_h(g)/d_h(s)}$,
as expected by dimensional reasons.

\vfill

\newpage

\section{Introduction}

Two-dimensional quantum gravity offers a unique chance to gain understanding
of the quantum aspects of space-time, while still using conventional 
techniques of quantum field theory. Of course the dynamics of gravitons 
cannot be addressed in such a theory, but the problem of defining and 
calculating reparameterization 
invariant observables still exists, and is even emphasized in 
two dimensions where each geometry appears with the same 
weight (apart from the cosmological constant term). From this point of view 
one is as far away from a classical picture of space-time as one 
can possibly be.

For pure two-dimensional 
quantum gravity the fractal structure of quantum space-time is by 
now well understood. In fact, many facets of this structure have simple 
analogies in the case of the free 
particle. Non-trivial aspects of fractal geometry associated with the 
propagation of the free particle refer to the {\it extrinsic} geometry of 
the world lines of the particle, as they appear in the path integral.
One of the most fundamental results of quantum physics, as compared 
to classical physics, is that a generic world line of the particle 
has extrinsic Hausdorff dimension $d_H=2$. The implications even for 
interacting quantum field theory are well known. Let us only mention the 
proof that there exists no interacting $\phi^4$ theory in dimensions 
larger than four, which is based on the fact that $d_H =2$.

In pure \tdqg\ we are instructed in the path integral formalism  to integrate 
over all space-time geometries\footnote{By geometries we mean metrics modulo 
reparameterization invariance.}.
One finds that the {\it intrinsic} Hausdorff dimension $d_h(g)$,
defined in the ensemble of space-time geometries, is four.

If a conformal field theory with central charge $c$ is coupled to \tdqg\
the situation is more complicated. By a study of the diffusion
equation in quantum Liouville theory the following formula for
$d_h(g,c)$ was derived~\cite{Kaw}:
\beq{w}
d_h(g,c)= 2\, \frac{\sqrt{25-c}+ \sqrt{49-c}}{\sqrt{25-c}+\sqrt{1-c}}.
\eeq
This formula gives $d_h=4$ for $c=0$ and there exist
extensive numerical simulations which have confirmed the formula for $c = -2$
with high precision.
However, an alternative formula for $d_h$ was derived in \cite{ik},
namely:
\beq{ik}
d_h(s,c) = -\frac{2}{\g(c)},~~~~\g(c) = \frac{c-1-\sqrt{(1-c)(25-c)}}{12}.
\eeq
Also this formula gives $d_h=4$ for $c=0$. But it differs from \rf{w}
for all other values of $c$. In particular  it
disagrees with \rf{w} for $c=-2$.
The analysis leading to \rf{w} is based on geometry defined in terms
of geodesic distance while the derivation of \rf{ik} uses the concept
of spin-clusters, as we will describe in more detail in the next section.
Thus it is natural to conjecture
that \rf{ik} does not describe the fractal structure of space-time
when measured in terms of the geodesic distance. Nevertheless, there
is scaling corresponding to \rf{ik}, as we will verify, 
and in fact this scaling is in many ways the one which
relates most naturally to the scaling known in matrix models.

In this article we  show that there  exists  scaling
of spin clusters and that the standard laws of scaling for spin clusters
predict $d_h(s,c{\rm=}1/2) = 6$ for \tdqg\ 
coupled to Ising spins at the critical
temperature. One can even measure $d_h(s,c{\rm =}1/2) \approx 6$
numerically using spin clusters,
while a similar measurement using the same size lattices  and
the same temperature results in $d_h(g, c{\rm=}1/2) \approx 4$ 
when the geodesic distance is used as a probe.

The rest of the article is organized as follows. In section 2 we review
the derivation of \rf{ik} using the representation of conformal field 
theories with central charge $c \in [-2,1]$ as $O(n)$ or loop gas
models. For the I
sing model this implies a representation in terms of
spin cluster boundaries. We furthermore show that the arguments which 
lead to~\rf{ik} imply that the concept of Hausdorff dimension breaks
down in the so-called dense phase of the $O(n)$ models which in Ising
model terminology corresponds to the unmagnetized phase.
In section 3 we present some simple scaling
arguments for spin clusters and explain how they lead to $d_h(s) = 16$
in flat space, while they give $d_h=6$ after coupling the Ising model
to \tdqg. In section 4 we show that one can indeed measure 
$d_h(s) \approx 6$ if one uses spin-cluster boundaries to define 
``distance'', while the use of geodesic distance leads to 
$d_h(g) \approx 4$. Finally section 5 contains a discussion.

\section{Loop gas representation and ``distance''}

The $O(n)$ model on a regular three-coordinate lattice can be given
an interpretation as a loop gas model. Expanding the exponential of
its action, truncating the series after the first non-trivial term and
summing over spin variables, the only terms which survive are
those to which one can associate a collection of closed,
self-avoiding and non-intersecting loops living on the 
lattice~\cite{DMNS81}. The $O(1)$ model, i.e.\ the Ising model 
has the special property
that the above mentioned truncation produces a model which after a suitable
redefinition of its coupling constant can be identified as the original
model. If we consider a regular three-coordinate lattice
 with the topology of the
sphere the partition function of the model can be written
as\footnote{A three-coordinate lattice of spherical topology
can of course not be completely regular. Some defects must be introduced.} 
\beq{2.2}
Z_T(K) = \sum_{\{\cL\}} \frac{1}{C_T(\{\cL\})}\; K^{L(\cL)}\,
n^{\cal N(\cL)}
\eeq
where the sum is over loop configurations, $\{\cal L\}$, having the
above mentioned properties. The coupling constant $K$ 
encodes the temperature dependence,
${\cal N}( \cL)$ is the number
of loops in $\{ \cL\}$ and $L( \cL)$ is the total length of the
loops in $\{ \cL\}$. Finally, $C_T(\{ \cL\})$ denotes the order of the
automorphism group of the lattice $T$ with the loop
configuration $\{L\}$. 
The representation~\rf{2.2} for the partition
function makes it possible to extend the definition of the model to
non-integer and negative values of $n$ by analytical continuation. It
is well-known that for $n\in [-2,2]$ the $O(n)$ model has a second
order phase transition at some critical point
$K=K_c(n)$~\cite{Nie82}. The continuum theory which can be defined at 
this critical point can be identified as a conformal field theory
with the central charge, $c$, depending on $n$ in the following way~\cite{DF84}
\beq{2.1}
c=1-6\,\frac{\n^2}{1+\n},~~~~~~~n = 2 \cos (\n \pi),~~~~~\n \in [0,1].
\eeq
In particular if $\nu=\frac{1}{m}$ one gets the minimal unitary model of
the type $(m,m+1)$. The $O(n)$ model is also critical for
$K<K_c(n)$. This part of the coupling constant space is known as 
the dense phase of the model, the name
referring to the fact that in this phase  
only a vanishing fraction of the lattice is occupied by loops. 
At $K=K_c(n)$ the condensation of loops
ceases and the model is said to be in the dilute phase.

By duality we can view the $O(n)$ model described above as being defined on
a regular triangulation and with that interpretation the model
can be coupled to two-dimensional quantum gravity by the standard recipe. 
This leads to the so-called $O(n)$ model on a random lattice whose partition 
function reads~\cite{Kos89,DK88}
\beq{2.3}
Z(\m,K) = \sum_{T\sim S^2} \e^{-\m N_T} Z_T(K).
\eeq
Here we sum over all triangulations with the topology of the sphere and
it is understood that the loops live on the lattice dual to the triangulation.
The quantity
$N_T$ is the number of triangles in the triangulation and $\mu$ is the
cosmological constant. For $n=1$ the model~\rf{2.3} is
exactly equivalent to the Ising model on a random lattice (with no
magnetic field). The spins reside on the vertices of the triangulation
and the loops are spin boundaries separating regions with spin $+$
from regions with spin $-$. Let us denote the triangles traversed by
loops as decorated triangles and those not traversed by loops as
non-decorated triangles. Then we can also write the partition function
as
\beq{kappa}
Z(\m,K) = \sum_{T\sim S^2} \e^{-\m N_{nd}} 
\sum_{\{ {\cal L} \}}\frac{1}{C_T(\{ \cL\})}\,\kappa^{N_d}\, n^{{\cal
N}(\cL)}, \hspace{1.0cm} \kappa=e^{-\mu}\,K 
\eeq
where $N_{nd}$ is the number of non-decorated triangles and $N_d$ is
the number of decorated triangles. 
In the coupling constant space of
the model~\rf{kappa} there exists a line of critical points 
beyond which the partition function does not exist. On this critical line
there is a particular point $(\kappa^*,\mu^*)$  at which a phase transition 
takes place. This phase transition is the analogue of the phase 
transition at $K=K_c(n)$ seen on a regular lattice. For $\kappa >\kappa^*$
the singular behavior of the partition function is due to the radius of
convergence in $\mu$ being reached while for $\kappa<\kappa^*$ the singular
behavior of the partition function is due to the radius of convergence
in $\kappa$ being reached. If we approach the critical line in the region
where $\kappa>\kappa^*$ by letting $\mu\rightarrow \mu_c(\kappa)$, the 
average number of non-decorated triangles diverges while the average number 
of decorated triangles stays finite. Not surprisingly, the resulting continuum
theory belongs to the universality class of pure gravity. If, on the other
hand, we approach the critical line in the region where $\kappa<\kappa^*$
by letting $\kappa\rightarrow \kappa_c(\mu)$ the average number of decorated 
triangles  diverges while the average number of non-decorated triangles
stays finite. The model is hence in its dense phase.
 Finally, if we let $(\mu,\kappa)\rightarrow (\mu^*,\kappa^*)$ both the 
number of decorated and the number of
non-decorated triangles diverge but the decorated triangles fill a
vanishing fraction of the surface in the scaling limit. This is
 the dilute phase of the model and for $n\in[-2,2]$ the scaling
behavior of the corresponding continuum theory can be identified as
that of a conformal field theory of the type described
in~\rf{2.1} dressed by quantum gravity~\cite{Kos89,DK88}. 
In the following we shall be
particularly interested in the case $n=1$, $\nu=\frac{1}{3}$,
i.e.\ the Ising model on a random lattice. For the Ising model the
dense phase ($\kappa<\kappa^*$) corresponds to the unmagnetized
phase whereas the dilute phase $\kappa=\kappa^*$ corresponds to the
phase transition point. The third case $\kappa>\kappa^*$ corresponds
to the situation where the model is completely magnetized.

In order to study the fractal properties of space-time in the case
where Ising spins are coupled to 2D quantum gravity it is convenient
to introduce the loop-loop correlation function. The loop-loop
correlation function is the amplitude for (triangulated) manifolds
with the topology of the cylinder where one boundary is denoted as the
entrance loop and the other one as the exit loop and where the two
boundaries are separated by a certain given distance. In the case of
the Ising model we shall restrict ourselves to the situation where all
spins along a given boundary are identical but where the direction of
the spins is not necessarily the same for the entrance and the exit
loop. 

For a given triangulation and a given spin configuration we
define a discrete distance or time variable, $t_D$, referring to loops
of the above mentioned type, as follows. We start from a loop along
which the spins are aligned; the entrance loop. All triangles which
share a link with the entrance loop have a distance one from this
loop. In case a spin boundary passes through one of these triangles
all the other triangles through which the spin boundary passes are likewise
said to have distance one from the entrance loop. Deleting the
entrance loop and all triangles assigned the distance one creates a
new boundary, in general consisting of several connected
components. Along each of these the spins are aligned but on only one
of them the direction of the spins is the same as on the original
loop. One can continue this process and each of the connected
components of the boundary obtained after $t_D$ steps is said to have
distance $t_D$ to the entrance loop. It is obvious that our definition
of the distance variable relies only on the loop configuration 
$\{ \cL \}$. Hence such a distance variable may be defined for any $n$.

The decomposition of the surface
that we have described here is known as slicing. There exists a
closely related procedure, known as peeling, which has the advantage
that the time variable introduced, $t_s$, is a continuous variable.
We refer to~\cite{Wat95,AKW97} for details. The
two versions of the time variable, $t_D$ and $t_s$, are believed to
agree in the scaling limit. Using the peeling decomposition one can
derive a differential equation for the loop-loop
correlator~\cite{AKW97}. 
For simplicity, let us give the differential equation in a symmetrized
form
\beq{2.4}
\frac{\partial }{\partial t_s} G_-(p,q,t_s)=
-\frac{\prt}{\prt p}\left\{(2+n)W_{s+}(p)G_{+}(p,q,t_s)
+(2-n)W_{s-}(p)G_-(p,q,t_s)\right\}.
\eeq
Here $G(p,q,t_s)$ denotes the (discrete) Laplace
transform of the loop-loop correlator $G(l_{en},l_{ex},t_s)$ where
$l_{en}$ and $l_{ex}$ are the (discrete) lengths of the entrance and the
exit loop respectively and $t_s$ the distance between the two loops.
The function $W(p)$ is the amplitude for a disk along the boundary of
which the spins are aligned.
The subscripts $\pm$ on a given function refer to its even and odd
part in the following sense
\beq{evenodd}
 f_{\pm}(p)=f(p)\pm f(\frac{1}{\kappa}-p).
\eeq
(In the case of $G(p,q,t_s)$ it is understood that the
transformation~\rf{evenodd} should be applied only to its first argument.)
Finally, the subscript $s$ on $W(p)$ refers to the singular part. For
details, see for instance reference~\cite{EK95} where an explicit
expression for $W_s(p)$, valid for any $n$, was written down.

Let us now consider the continuum limit of the loop gas model and let
us restrict ourselves to the case $n\in [-2,2]$. To define a continuum
theory corresponding to the {\it dilute} phase of the $O(n)$ model on
a random lattice we must scale $\mu$ and $\kappa$ to their critical
values $\mu^*$ and $\kappa^*$ in the following way~\cite{Kos89,DK88}
\beq{2.6}
\m -\m^* \sim \ep^2 \L,~~~~~\k-\k^* \sim  \ep \L,
\eeq
where $\L$ is the continuum cosmological constant and $\epsilon^2$ a
parameter with the dimension of volume. In addition, the variable $p$
conjugate to the loop length, must be scaled as
\beq{2.7}
p-p_c = p-\frac{1}{2\k^*} \sim \ep \sg,
\eeq
$\sg$ being the continuum variable conjugate to the continuum loop
length. Under these circumstances the singular part of the disk
amplitude behaves as
\beq{disk}
W_{s\pm}(p)\sim \epsilon^{1+\nu}W_{s\pm}(\sigma).
\eeq
It now follows that the differential equation~\rf{2.4} has a
non-trivial continuum limit only if the time variable, $t_s$, scales as
\beq{scalt}
t_s=\epsilon^{-\nu}\tau 
\eeq
where $\tau$ is some continuum time variable. Combining~\rf{2.6}
and~\rf{scalt} we now see that $\tau$ has the dimension of
(volume)$^{\nu/2}$, i.e. the fractal dimension of space time, when
based on $\tau$ as a measure of distance, would be $d_h=2/\nu$. In
particular, for the minimal unitary model of type $(m,m+1)$, which 
has central charge $c(m)=1-6\frac{1}{m(m+1)}$ and corresponds to 
$\nu=\frac{1}{m}$, we get
\beq{dh}
 d_h(s,c(m))=2m.
\eeq
Let us now apply the same arguments to the dense phase of the $O(n)$
model. We remind the reader that in Ising model terminology the dense
phase is the unmagnetized phase. In this case the continuum
cosmological constant must be associated with the coupling constant
$\kappa$, i.e.
\beq{scalk}
\kappa-\kappa_c\sim\epsilon^2 \Lambda
\eeq
whereas the scaling to be imposed on the variable $p$ is again 
$p-p_c\sim \epsilon \sigma$. 
Now the singular part of the disk amplitude behaves as
\beq{Ws}
W_{s\pm}(p)\sim \epsilon^{1-\nu}W_{s\pm}(\sigma)
\eeq
Hence we are led to the conclusion that the time variable $t_s$ must
scale as $t_s\sim \epsilon^{\nu}\tau$. This means that the linear
extent of our manifolds scales as a negative power of the area. The
concept of Hausdorff dimension simply breaks down!

The above presented arguments 
leading to $d_h(s,c(m)) =2m$ have the virtue that 
they clearly allow an identification of the ``time'' variable used, 
also at a discretized level, at least in the case of the Ising model ($n=1$)
where the equivalence with the loop gas model is exact and where
no analytic continuation in $n$ is needed.
Thus one can directly use 
the above definition of ``time'' or ``distance'' in numerical simulations 
and one should be able to measure $d_h(s,c{\rm=}1/2) =6$ at the phase
transition point.
We will show that numerical simulations support
$d_h(s,c{\rm=}1/2) =6$.
Similarly, one should be able to see that the Hausdorff dimension
$d_h(s)$ becomes meaningless in the unmagnetized
phase and we do see this in our simulations.
It is clear that there is no obvious reason to expect the definition 
of ``time'' given above to be related to more conventional definitions
like the one based on geodesic distance. {\it A priori}
 one cannot 
rule out the possibility that the two distances
 will be related in the scaling limit but we
 will show that it is unlikely that there exists a simple relation.

A similar conclusion was reached for a $c=-2$ theory coupled to 
\tdqg\ in a recent paper~\cite{AKW97}. For $c=-2$ it is possible to carry the 
calculation of $G(p,q,t_s)$ much further and to calculate the so-called 
two-point function in almost the same detail as for pure gravity.
In this way one can study the fractal structure based on the ``time''
$t_s$ in more detail and does not have to rely entirely
 on dimensional arguments like   
\rf{Ws}. When one uses the loop gas ``time''
variable as defined above, one finds $d_h(s,c=-2)=2$
in agreement with \rf{ik}. 
At the same time it is possible to perform 
very precise numerical measurements of $d_h(g,c{\rm=-}2)$ based on 
geodesic distance, due to the fact that 
one can avoid Monte Carlo simulations and use instead recursive sampling
of configurations. The result is a convincing agreement with formula 
\rf{w} and a disagreement with \rf{ik}.  In the case $c=-2$ is not possible 
to measure directly the fractal dimension as defined 
by the loop gas ``time'', since $c=-2$ corresponds to an analytic 
continuation of the $O(n)$ model 
to $n=-2$. In the case of the Ising model we do not have 
the analytic expression for the two-point function
but we can, as mentioned, measure $d_h(s)$ in numerical 
simulations.

\section{Scaling of spin clusters and fractal dimension}

It is possible to derive $d_h(s,c{\rm=}1/2)=6$ from the scaling properties 
of the Ising spin clusters. Let us first present the arguments
in flat space, i.e.\ we consider a regular triangular 
lattice with the topology of the sphere\footnote{A triangulation of the
sphere can of course not be completely regular. Some vertices necessarily
have an order different from six.}.

Let us recall some basic properties of two-dimensional spin clusters.
Consider a system of linear size $L$ and volume $V \sim L^d$, $d=2$. 
In the context of finite size scaling we have at the pseudo-critical 
point, where the correlation length $\xi \sim L$, that the absolute 
value of the magnetization per volume, $m$, is given by
\beq{3.1}
|m| \sim  L^{-\b/\n} \sim V^{-\b/\n d},
\eeq
where $\n$ is the critical exponent for the spin-spin correlation length.
The absolute value of the total magnetization is thus given by 
\beq{3.2}
|M| = |m| V \sim V^{1-\b/ \n d}.
\eeq
The magnetization in  the two-dimensional Ising model is   
essentially determined by the largest spin cluster. If $S$ denotes
the average maximum size of such a spin cluster we have 
\beq{3.3}
|M| \sim S.
\eeq
At the same time the average maximum spin cluster fills out a fraction of the 
whole lattice, in this way defining a fractal dimension $D$ of the 
spin cluster:
\beq{3.4}
S \sim L^D \sim V^{D/d}.
\eeq
>From \rf{3.2}-\rf{3.4} we conclude that 
\beq{3.5}
\frac{D}{d} = 1- \frac{\b}{\n d}~~~~( = \frac{15}{16}~\mbox{for the 
Ising model}).
\eeq
The only additional thing we need to know is that the fractal dimension of 
the spin clusters is equal to the fractal dimension of their boundaries,
a generic phenomenon in percolation theory.

On the regular triangulated lattice we can now define a fractal 
dimension as follows, using the distance in terms of spin boundaries
defined above. Since the largest spin boundaries  dominate compared to 
ordinary boundaries on the lattice (the largest of them has dimension 
15/8, compared to 1 for an ordinary, non-fractal  boundary) the ``filling''
of the complete volume, starting from a single spin, will be determined 
by the large spin clusters. let $R$ denote the number of ``radial steps''
needed for such a filling. Symbolically we can write:
\beq{3.6}
V \sim  V^{D/d} \times R,
\eeq
i.e.
\beq{3.7}
{\rm dim}\; V^{1/d_h(s)} \equiv {\rm dim}\; R = {\rm dim}\; V^{\b/\n d},
~~~~{\rm or}~~d_h(s) = \frac{\n d}{\b}.
\eeq
For a regular triangulation this mean field argument leads to $d_h(s) =16$,
which should be compared to the ``fractal'' dimension $d_h=2$ defined by
using ordinary ``geodesic'' distance on the lattice!

The virtue of the above arguments is that they can be repeated word 
by word in the case of the Ising model coupled to \tdqg\footnote{The only 
point which is slightly unclear is how to write $L^d = V$. Precisely which 
$d$ should be used: $d=2$, or $d_h$ as determined by geodesic distance, or 
$d_h$ determined by the loop gas ``time''?. We do not have to choose since 
only the product $\n d$ appears, and it can be calculated unambiguously 
from the KPZ relations.}. From the KPZ scaling relations we get 
\beq{3.8}
\left. \frac{\b}{\n d} \right|_{\rm gravity-dressed} = \frac{1}{6},
\eeq
i.e.\ the fractal dimension is $d_h(s) =6$ according to 
the loop gas definition 
of ``time''. Intuitively it seems correct that the fractal dimension defined
by means of spin boundaries should decrease compared to the similar 
fractal dimension in flat space since the spin clusters are smaller 
in the Ising model coupled to 2d gravity than in the Ising model in flat 
space. The reason for this is that the fluctuating geometry with many baby
universe out-growths makes it difficult to create large spin clusters.

\section{Numerical verification}

Numerical simulations of the Ising model coupled to 
\tdqg\ using the geodesic distance have so far 
failed to reproduce the prediction $d_h =6$ discussed in the last 
two sections. Rather, the measurements favor 
$d_h \approx 4$ (see \cite{syracuse,ajw,aa}). 
It has been argued that one cannot measure $d_h = 6$ because it 
would require very large lattices \cite{dass}. Other arguments in disfavor of 
practical measurements of $d_h=6$ have also been presented \cite{mark}.
Below we will present arguments which show that one gets a clear 
indication of  $d_h \approx6$ even on quite small lattices, provided one uses
as ``time'' the parameter which was actually used to derive $d_h=6$,
namely the loop gas ``time'' $t_s$ (or more precisely $t_D$), defined above.

The numerical experiments performed so far have concentrated on measurements
of the fractal dimension by means of the volumes $n_N (t_g)$ 
of spherical shells of geodesic radius $t_g$. 
If the universes are of total volume 
$N$ (the number of triangles in the triangulation) the functional form 
of $n_N(t_g)$ is
\beq{4.1}
n_N (t_g) = N^{1-1/d_h(g)} F_g(x_g),~~~~x_g=\frac{t_g}{N^{1/d_h(g)}}.
\eeq
This relation is basically a finite size scaling relation which 
has been proven analytically for $c=0$, 
and was found very powerful in numerical simulations
also for $c \neq 0$.  It has been verified in
great detail for $c=-2$ where, as already mentioned,
 recursive sampling of configurations
allows for better statistics and larger triangulations than
 the standard Monte Carlo techniques \cite{many,many1}.

\begin{figure}[htb]
\vspace{-4.5cm}
\centerline{\hbox{\psfig{figure=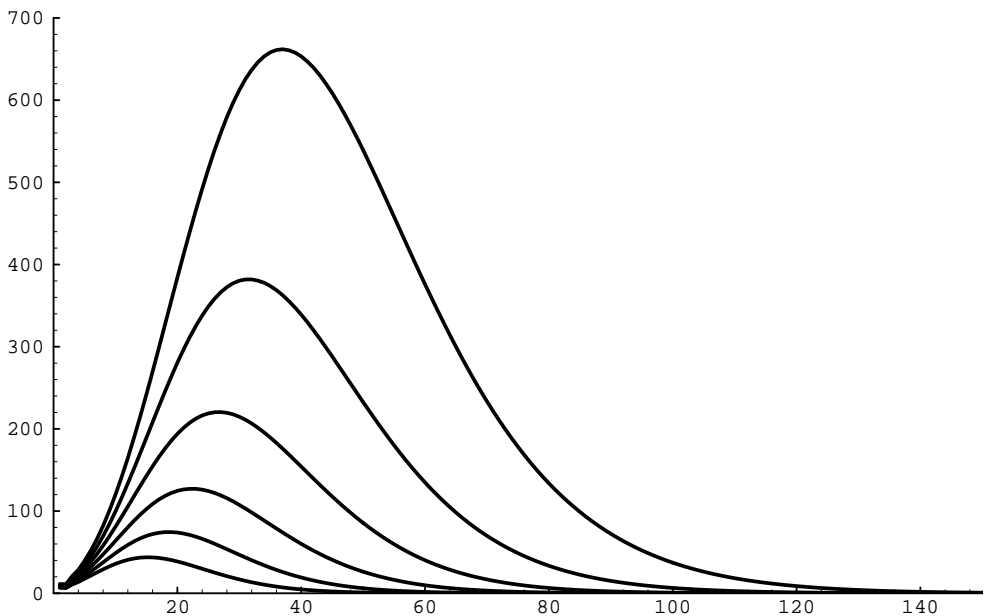,height=18cm,angle=0}}}
\vspace{-5cm}
\caption[fig1]{The distributions $n_N(s,t_D)$ for $N= 1K,2K,4K,\dots,32K$.}
\label{fig1}
\end{figure}

We can perform the same measurements using $t_D$ as the ``radial'' distance 
out to ``spherical'' shells. It is then natural to expect scaling relations
of the same kind as \rf{4.1}:
\beq{4.4}
n_N(s,t_D) = N^{1-1/d_h(s)} F_s(x_s),~~~~~x_s = \frac{t_D}{N^{1/d_h(s)}}.
\eeq   
In fig.\ \ref{fig1} we show the result of the measurement of the 
distributions $n_N(s,t_D)$ for $N=1K,2K,4K,\dots,32K$.
The scaling given by \rf{4.4} should hold in the $N \to \infty$ limit. For
finite $N$ one expects it to be modified by various finite size corrections.
One typical effect was already observed in the cases $c=0$ and $c=-2$ 
\cite{ajw,many,many1} and an analytic justification  in the case of $c=0$ 
can be found in \cite{many1}. It
consists of a finite redefinition of the scaling variable $x_s$ by a ``shift''
$\epsilon$:
\beq{4.5}
x_s = \frac{t_D - \epsilon}{N^{1/d_h(s)}}.
\eeq
Another expected effect is in fact more standard: on a finite lattice a
definition of the pseudo-critical point $\kappa_c(N)$ 
is not unique. Usually it is defined
by the position of the maximum of the specific heat and as a rule it approaches
the real critical point only when $N \to \infty$ like  
\beq{4.6}
\kappa_c(N) \approx \kappa^* + const.\cdot \frac{1}{N^{\alpha}}.
\eeq
For different definitions of the pseudo-critical point the constant is usually
different and we do not know a priori what definition one should take to
ensure the correct measurement of the Hausdorff dimension $d_h(s)$. In the
case of the Ising model we are, however, in a comfortable situation, because
we know the exact value of the critical coupling, namely  
$\b^*=-\oh \log \kappa^* =\oh \ln (1+1/\sqrt{7})\approx 0.1603$ 
\cite{boulatov}. We can
therefore approach the $N \to \infty$ limit at the fixed value  $\kappa = 
\kappa^*$. For finite $N$ the system will not in general be  critical and
we expect therefore that the measured ``effective'' Hausdorff dimension
will be a function of the volume $N$. How large $N$'s one has to 
use in order to get a good approximation to $d_h(s)$ depends one $d_h(s)$
itself. A larger value of  $d_h(s)$ will in general require larger values
of $N$ since one expects that $N$ should be sufficiently large to allow
the inequality
\beq{4.6a}
1 \ll t_D \ll N^{1/d_h(s)}.
\eeq
Thus we expect that one might have to go to quite large $N$ precisely
at $\kappa^*$, an expectation which is supported by the fact that 
the spin boundaries which are used to step out in ``radial'' direction
grow at the critical point.

In the following we shall measure the ``effective'' Hausdorff dimension
$D_h(N,\kappa)$ by comparing the scaling functions $n_N(s,t_D,\kappa)$
for three values of volume: $N/2$, $N$ and $2N$, where $N=2K,4K,8K$ and $16K$. 
In practice we use two
estimates coming from the scaling of the maximum of the distribution
($\propto N^{1-1/D_h}$) and of the dispersion ($\propto N^{2/D_h}$),
respectively.
Notice that these two quantities are not sensitive to the finite shift
mentioned above. The difference between the two estimates can be used as
a measure of the error.  

\begin{figure}[htb]
\vspace{-4.5cm}
\centerline{\hbox{\psfig{figure=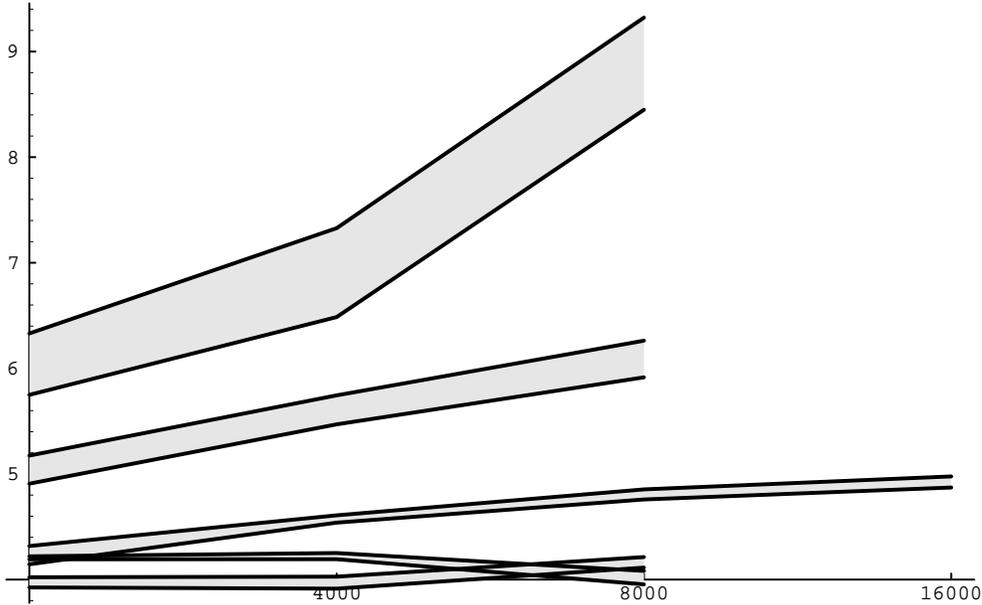,height=18cm,angle=0}}}
\vspace{-5cm}
\caption[fig2]{$D_h(N,\kappa)$ shown as a function of $N$ 
for (from top to bottom) $\b = -\oh \log \kappa =0.14,0.15$, ($N=2K,4K$ and
$8K$), 
$\b^* \approx 0.1603$ ($N=2K,4K,8K$ and $16K$) and 
$\b =0.17,0.18$ ($N=2K,4K$ and $8K$).
}
\label{fig2}
\end{figure}

The results of the measurements of $D_h(N,\kappa)$ as a function of $N$ 
for $\kappa = 0.180$, $0.170$,
$\kappa^*$, $0.150$ and $0.140$ are presented in fig.\ \ref{fig2}.
 We see that
for $\kappa > \kappa^*$ the value of $D_h(N,\kappa)$ 
is consistent with $d_h(s) = 4$, which
is the value for pure gravity ($\kappa \to \infty$). For $\kappa < \kappa^*$
the functions $D_h(N,\kappa)$ grow with $N$ at least logarithmically, 
which suggests that $D_h(N,\kappa) \to \infty$ for $N \to \infty$
for these values of $\kappa$. 
Exactly at the critical point $\kappa=\kappa^*$ 
we see a different behavior of $D_h(N,\kappa)$,
consistent with $D_h(N,\kappa^*) \to 6$ for $N \to \infty$.
It is also clear from the slow rise of $D_h(N,\kappa^*)$ that it  
will require much larger values of $N$ if one wants to obtain a 
high precision measurement of $\lim_{N \to \infty} D_h(N,\kappa^*)$.

\section{Discussion}

For the Ising model we have verified that the two-point function
defined as a function of the loop gas ``time'' $t_s$ indeed
scales with a $d_h(s)  \approx 6$, i.e.\ in agreement with \rf{ik}.
On the other hand the use of geodesic distance yields $d_h(g) \approx 4$.
The results using geodesic distance are in perfect agreement with
\rf{w} for $c=-2$. However, for the Ising model coupled to
\tdqg\ the result $d_h(g) \approx 4$ are only marginally in agreement with
the prediction $d_h =4.2...$ by \rf{w}. But certainly \rf{ik} is ruled out,
when geodesic distance is used. The same conclusion was reached for
$c=-2$ coupled to \tdqg\ \cite{many,many1}. 

While the loop gas time $t_s$ lacks any obvious geometric interpretation
except for $c=0$, where it coincides with the ordinary geodesic distance,
it is nevertheless the scaling governed by the dimensionality of $t_s$
which appears in the simplest way in matrix model calculations.
Further, it is for this choice of ``time'' variable that it is
simplest to develop a string field theory and a transfer matrix 
formalism. It follows from the transfer matrix formalism that 
the two-point function for fixed cosmological constant will decay
exponentially in $t_s$~\cite{AKW97}. By an inverse Laplace transformation this 
implies the following large $t_s$ behavior for fixed space-time 
volume $N$:
\beq{5.1}
n_N(s,t_s) \sim \exp\left\{
-\left(\frac{t_s}{N^{1/d_h(s)}}
\right)^{\frac{d_h(s)}{d_h(s)-1}}\right\}.
\eeq
>From dimensional arguments it is tempting to relate $t_s$ and $t_g$ as follows
\beq{5.2}
t_s^{d_h(s)} \sim t_g^{d_h(g)}..
\eeq
This suggests that the long distance behavior of $n_N(g,t_g )$ should be
\beq{5.3}
n_N(g,t_g) \sim \exp\left\{-\left(
\frac{t_g^{\frac{d_h(g)}{d_h(s)}}}{N^{1/d_h(s)}}
\right)^{\frac{d_h(s)}{d_h(s)-1}}\right\}.
\eeq
This prediction can be tested numerically by a study of the tail of the 
distribution $n_N(g,t_g)$. For $c=-2$ we have a power of $t_g^{3.56}$
(compared to a naive power $t_g^{1.39}$, assuming exponential decay 
of the two-point function for fixed cosmological constant, using
geodesic distance),
while for $c=1/2$ the power is $t_g^{4/5}$ (compared to $t_g^{4/3}$).
These predictions seem {\it not} to be satisfied numerically
and it is unlikely that there is a simple relation like \rf{5.2}
between $t_s$ and $t_g$.

\end{document}